\documentclass[aps,prb,reprint,showpacs,superscriptaddress]{revtex4-1}
\usepackage[dvipdfm]{graphicx}

\begin{document}


\title{Josephson effects in one-dimensional supersolids}


\author{Masaya Kunimi}
\affiliation{Department of Basic Science, The University of Tokyo, Tokyo 153-8902, Japan}
\author{Yuki Nagai}
\affiliation{CCSE, Japan Atomic Energy Agency, 5-1-5 Kashiwanoha, Kashiwa, Chiba 277-8587, Japan}
\affiliation{CREST(JST), 4-1-8 Honcho, Kawaguchi, Saitama, 332-0012, Japan}
\author{Yusuke Kato}
\affiliation{Department of Basic Science, The University of Tokyo, Tokyo 153-8902, Japan}

\date{\today}

\begin{abstract}

We demonstrate that superflow past an obstacle is possible in a solid phase in the one-dimensional Gross-Pitaevskii equation with a finite-range two-body interaction. The phenomenon we find is analogous to the DC Josephson effect in superconductors and we deduce the ^^ ^^ Josephson relation " between the current and phase difference of the condensates separated by the obstacle. 
We also discuss persistent current and nonclassical rotational inertia in annular container with a penetrable potential barrier.
The phase diagram in the plane of the current and the interaction strength is given.  
Our result provides a simple theoretical example of supersolidity {\it in the presence of an obstacle}.
\end{abstract}


\maketitle

\section{Introduction}\label{sec:intro}
Supersolid is the quantum solid with a superflow property.
Although the possibility of supersolid has been discussed theoretically about 40 years ago\cite{Andreev1969,Chester1970,Leggett1970}, extensive and intensive studies have started since Kim and Chan have reported\cite{Kim2004_N,Kim2004_S,Kim2005,Kim2006} the first evidence for the non-classical rotational inertia\cite{Leggett1970} in a solid phase of ${}^4$He in the torsional oscillator experiments. 
Subsequently, experimental studies have revealed various puzzling nature in supersolid $^4$He\cite{Prokofev2007,Balibar2008,Galli2008,Balibar2010}: enhancement of elasticity having the same temperature dependence as the resonance frequency shift\cite{Day2007}, the sensitivity of resonance frequency shift on history of annealing process\cite{Rittner2006} and small concentration of $^3$He\cite{Day2007}, superflow along the grain boundaries\cite{Sasaki2006}, ultraslow relaxation dynamics in the dissipation and resonance frequency shift in the torsional oscillator experiments\cite{Hunt2009}. As consistent explanations of these experimental results, several theories have been proposed such as defect-mediated superfluidity mechanisms\cite{Pollet2007,Boninsegni2007}, a non-superfluid glass picture\cite{Nussinov2007,Andreev2007} and intrinsic supersolidity of defect-free crystal\cite{Anderson2007}. 

Supersolid is also related to other systems; experiments on $^4$He film realized on graphite suggest a possibility of two-dimensional supersolid\cite{Shibayama2009}. Bose-Einstein condensate(BEC) of $^{52}$Cr might be another possible candidate of the system with a supersolid phase, owing to the long-range dipole-dipole interactions\cite{Griesmaier2005,Goral2002,Saito2009,Danshita2010}. The possibility of supersolid of Rydberg atoms has been studied very recently\cite{Henkel2010,Cinti2010}. 
Thus, the properties of intrinsic supersolids should be clarified in more detail from broader contexts.

As a phenomenological model of supersolid, Pomeau and Rica\cite{Pomeau1994} have investigated a modified version of the Gross-Pitaevskii (GP) model\cite{Gross1961,Pitaevskii1961}, which has a two-body repulsive interaction not of a contact-type but with a finite-range. The liquid has the excitation energy with roton minimum for sufficiently large repulsive interaction (or sufficiently high density of particles). For larger repulsive interaction, the ground state acquires the positional order detected by the Bragg peaks. It has been shown that the solid exhibits the non-classical rotational inertia \cite{Pomeau1994,Josserand2007,Aftalion2007,Sepulveda2008} and quantized vortices\cite{Pomeau1994} under a uniform rotation. While the solid phase could be identified with a supersolid from these results, it was found that the dissipationless flow around an obstacle is not possible\cite{Pomeau1994}. It is thus important to see whether the supersolid can sustain the superflow in the presence of obstacles or not. This is the issue addressed in the present paper. 

As a simplest set-up, we consider the GP equation with a finite-range two-body interaction in one-dimensional systems. We show that the steady flow state exists in the solid phase even in the presence of an obstacle. The superflow in the presence of an obstacle in a one-dimensional system can be regarded as a DC Josephson current. That is, we examine superflow property of solid phase(, which we call supersolidity in this paper) through the DC Josephson effect\cite{Josephson1962}. The Josephson effects observed in ${}^4$He and in BEC in cold atoms serve an evidence for superfluidity\cite{Hoskinson2006,Levy2007}. 
%
In addition to the Josephson effect, we also discuss two other important superflow properties: the persistent-current effect and Hess-Fairbank effect in supersolid state in a ring with a potential barrier. The former effect is a flow property of metastable states while the latter is a property of thermodynamically stable state in which the whole or a part of supersolid is at rest against in a container rotating with a sufficiently small angular velocity. These two effects are considered as two fundamental properties of superfluids and correspond, respectively, to persistent electric current and Meissner effects in superconductors.

This paper is organized as follows. 
In Sec.~\ref{sec:model}, we introduce the GP equation with a finite-range two-body interaction and explain the method of numerical calculations. In Sec.~\ref{sec:results}, we first present the phase diagram in the plane of current $J$ and (dimensionless) two-body interaction strength $g$. We next present the results related to the Josephson effects in the solid phase. Further we discuss persistent current and present results on nonclassical rotational inertia in Hess-Fairbank effect. In Sec.~\ref{sec:discussion}, we discuss implications to systems in higher dimensions and possible difference of superfluid properties between the supersolid phase and conventional superfluids. The conclusion is given in Sec.~\ref{sec:conclusion}. 
\section{Model and Method}\label{sec:model}
We start with a one-dimensional GP equation with a finite-range interaction \cite{Pomeau1994}
\begin{eqnarray}
&&i\hbar\frac{\partial \Psi(x,t)}{\partial t}=-\frac{\hbar^2}{2m}\frac{\partial^2}{\partial x^2}\Psi(x,t)+\left[U(x)-\mu\right]\Psi(x,t)\nonumber \\
&&\quad \quad +\int^{L/2}_{-L/2}dy V(x-y)|\Psi(y,t)|^2\Psi(x,t)=0,
\label{eq:t-dep-GPequationforpsiform}
\end{eqnarray}
where $m$ is the atomic mass, $\Psi(x)$ is the condensate wave function. 
 $L$ is the system size. $V(x)$ and $U(x)$ represent, respectively, two-body repulsion and the potential barrier. 
$\mu$ denotes the chemical potential, which is determined by the condition on the total number of particles being\cite{note-1d}
\begin{eqnarray}
N=\int^{L/2}_{-L/2}dx|\Psi(x)|^2.\label{eq: chemical}
\end{eqnarray}
In the presence of potential barrier $U(x)$, there are two conservation laws\cite{Hakim1997}; one is equation of continuity of particle density
\begin{equation}
\frac{\partial |\Psi(x,t)|^2}{\partial t}+\frac{\partial J(x,t)}{\partial x}=0,
\label{eq: continuity}
\end{equation}
with the current\cite{note-1d}
\begin{equation}
J(x,t)=-\frac{i \hbar}{2m}\left[\Psi^*(x,t)\frac{\partial \Psi(x,t)}{\partial x}-\Psi(x,t)\frac{\partial \Psi^*(x,t)}{\partial x}\right]
\label{eq: Jxt}
\end{equation}
and the other is equation of continuity of local energy, the latter of which we do not discuss in this paper. 

If the stationary solution ($\partial \Psi/\partial t=0$) to (\ref{eq:t-dep-GPequationforpsiform}) exists, this solution has the current $J(x,t)$ being spatially constant (In the following, we denote by $J$ a constant current ). The stationary solution with finite $J$ represents a condensate with macroscopic dissipationless flow. The existence of such a stationary solution {\it in the presence of potential barrier $U(x)$} can be regarded as an evidence for superfluidity or supersolidity. In earlier works\cite{Hakim1997,Baratoff1970,Hakimnote,Baratoffnote}, stationary solutions with finite $J$ in the presence of a short-range barrier $U(x)\propto \delta(x)$ were found when two-body interaction is contact-type $V(x)\propto \delta(x)$. 
The critical current $J_{\rm c}$, above which stationary current-flow state is absent, was found to depend on the strength of barrier\cite{Hakim1997,Baratoff1970}. 

Time-independent GP equation is given by
\begin{eqnarray}
&&-\frac{\hbar^2}{2m}\frac{d^2}{d x^2}\Psi(x)+\left[U(x)-\mu\right]\Psi(x)\nonumber \\
&&\quad \quad +\int^{L/2}_{-L/2}dy V(x-y)|\Psi(y)|^2\Psi(x)=0.\label{eq:GPequationforpsiform}
\end{eqnarray}
Substituting a time-independent function $\Psi(x)\equiv A(x)e^{i\varphi(x)}$ into eq.~(\ref{eq:GPequationforpsiform}), we obtain two equations corresponding to the real and the imaginary parts
\begin{eqnarray}
&&{\cal G}[A(x),\varphi(x)]\equiv-\frac{\hbar^2}{2m}\left\{\frac{d^2}{dx^2}A(x)-\left[\frac{d\varphi(x)}{dx}\right]^2A(x)\right\}\nonumber \\
&&+\left[U(x)-\mu\right]A(x)+\int^{L/2}_{-L/2}\hspace{-0.75em}dyV(x-y)A(y)^2A(x)=0,\label{eq:GPequationforamplitude1}\\
&&\quad \quad  \quad \quad \quad \frac{d}{dx}\left[\frac{\hbar}{m}A(x)^2\frac{d\varphi(x)}{dx}\right]=0,\label{eq:contiunuityequationfora}
\end{eqnarray}
where $A(x)$ is the amplitude of the condensate wave function and $\varphi(x)$ is the phase of the condensate wave function. From eq. (\ref{eq:contiunuityequationfora}), which is (\ref{eq: continuity}) for stationary state , it follows that \begin{eqnarray}
J=\frac{\hbar}{m}A(x)^2\frac{d\varphi(x)}{dx}={\rm const}.\label{continuouseq}
\end{eqnarray}
Substituting eq.~(\ref{continuouseq}) into eq.~(\ref{eq:GPequationforamplitude1}), we can eliminate $\varphi(x)$ and obtain the GP equation for the amplitude
\begin{eqnarray}
\mathcal{L}[A(x)]&\equiv &\left\{-\frac{\hbar^2}{2m}\frac{d^2}{dx^2}+\frac{m}{2}\frac{J^2}{A(x)^4}+U(x)-\mu\right.\nonumber\\
&&\left. \quad +\int^{L/2}_{-L/2}dy V(x-y) A(y)^2\right\}A(x)=0.\label{eq:GPequationforAphiform}
\
\end{eqnarray}

In this paper, we take three kinds of boundary conditions, as explained below. 

In \ref{sec: phase diagram} and \ref{sec: Josephson}, we use solutions to eq.~(\ref{eq:GPequationforAphiform}) under the boundary condition
\begin{equation}
A(x+L)=A(x). \label{eq: bc-A}
\end{equation}
Once we obtain the solution to (\ref{eq:GPequationforAphiform}) with (\ref{eq: bc-A}), the expression for $\varphi(x)$ follows:
\begin{eqnarray}
\varphi(x)=\frac{mJ}{\hbar}\int^x_0dy\frac{1}{A(y)^2}.
\label{eq: varphi-A}
\end{eqnarray}
Here we fix overall phase factor such that $\varphi(x=0)=0$. The values of $\varphi(\pm L/2)$ at the boundaries, which are determined by $A(x)$ and $J$ through (\ref{eq: varphi-A}). 
This set-up describes the system attaching a current generator (one can regard  this one-dimensional system as an orifice attaching two reservoirs of superfluids or supersolids; See e.g. sec.~III of Ref.~\onlinecite{Anderson1966}.). On the basis of solutions to (\ref{eq:GPequationforAphiform}) under the boundary condition (\ref{eq: bc-A}), we present the results on the phase diagram in \ref{sec: phase diagram} and Josephson effect in \ref{sec: Josephson}. 

In \ref{sec: persistent}, we discuss persistent current in a ring with perimeter $L$. In this case, the boundary condition 
\begin{equation}
\Psi(x+L)=\Psi(x) \label{eq: bc-Psi}
\end{equation}
for (\ref{eq:GPequationforpsiform}) is used. 

In \ref{sec: Hess}, we discuss nonclassical rotational inertia in Hess-Fairbank effect in a ring rotating with angular velocity $\omega=2\pi v/L$, with use of the solutions to (\ref{eq:GPequationforpsiform}) satisfying the boundary condition
\begin{equation}
\Psi(x+L)=\Psi(x)e^{-i mvL/\hbar}. \label{eq: bc-Psi-modified}
\end{equation}
We can obtain the solutions to eq.~(\ref{eq:GPequationforpsiform}) under (\ref{eq: bc-Psi}) or (\ref{eq: bc-Psi-modified}) from the solutions to (\ref{eq:GPequationforAphiform}) satisfying (\ref{eq: bc-A}) by restricting the values of $J$. For the moment, we thus consider the latter solutions
.


For consideration of stability of solutions to (\ref{eq:GPequationforAphiform}) with (\ref{eq: bc-A}), it is helpful to note, as shown in appendix \ref{sec:appendixA}, that (\ref{eq:GPequationforamplitude1}) and (\ref{eq:contiunuityequationfora}) result from the stationary conditions of the following functional 
\begin{eqnarray}
K&\equiv& E-\hbar J\left[\varphi(L/2)-\varphi(-L/2)\right],\label{eq:energywithcurrent} \\
E&\equiv&\frac{\hbar^2}{2m}\int^{L/2}_{-L/2}dx\left|\frac{d\Psi(x)}{dx}\right|^2+\int^{L/2}_{-L/2}dxU(x)|\Psi(x)|^2\nonumber \\
&+&\frac{1}{2}\int^{L/2}_{-L/2}dx\int^{L/2}_{-L/2}dyV(x-y)|\Psi(y)|^2|\Psi(x)|^2
\end{eqnarray}
under the constraint (\ref{eq: chemical}). 
%
Thus the functional $K$ plays the role of ^^ ^^ energy" under the fixed value of $J$. The second term in eq.~(\ref{eq:energywithcurrent}) is inherent to the system with constant flow as remarked in ref.~\onlinecite{Hakim1997}. 

In the following, we set the potential barrier $U(x)$ to be 
\begin{eqnarray}
U(x)\equiv U_0\theta(d/2-|x|),\label{eq: U-def}
\end{eqnarray}
where $\theta(x)$ denotes the Heaviside step function. 
$V(x-y)$ denotes the soft core two-body interaction:\cite{Pomeau1994,Sepulveda2008}
\begin{eqnarray}
V(x-y)\equiv V_0\theta(a-|x-y|),
\end{eqnarray}
where $a$ is the interaction range. 

Let us measure the length, time, energy, $\Psi(x)$, and current in units of $a$, $\tau_0\equiv ma^2/\hbar$, $\epsilon_0\equiv \hbar^2/ma^2$,  $\sqrt{n_0}\equiv \sqrt{N/L}$, $J_0\equiv n_0\hbar/ma$ respectively. 
We define a dimensionless parameter $g$:\cite{Sepulveda2008}
\begin{eqnarray}
g\equiv \frac{2n_0ma^3V_0}{\hbar^2}, 
\end{eqnarray}
which characterizes the strength of interaction.

We solve eq.~(\ref{eq:GPequationforAphiform}) in the following procedures: First we consider a relaxation dynamics
\begin{equation}
\frac{\partial A(x,\tau)}{\partial \tau}=-{\cal L}[A(x,\tau)].
\label{eq: relaxation}
\end{equation}
The stationary solutions of eq.~(\ref{eq: relaxation}) are the solutions of eq.~(\ref{eq:GPequationforAphiform}). 
We discretize $x$ and $\tau$ with the backward-Euler method and then obtain a constituent non-linear equation for stationary solutions of eq.~(\ref{eq: relaxation}). The resultant equation is solved by the Newton method. We perform this calculation under a value of $\mu$ and repeat it until $\mu$ satisfies (\ref{eq: chemical}). When we obtain different solutions starting with different initial conditions for the Newton method, we choose the solution with the lower value of energy functional $K$ as the lowest energy state for given $J$. We take the system size as $L/a=40, 50$. In the following, we will present the results for $L/a=50$, which are almost similar to those for $L/a=40$. The number of the mesh points in real space is taken as $N_{\rm m}=2^{13}, 2^{14}$ or $2^{15}$. We take the range of the barrier $d$ to be $d/a=0.5,1, 1.4, 2$. We report the results for $d/a=1$ in the next section and relegated those for $d/a=0.5, 1.4$, and $2$ to Appendix \ref{sec:appendixB}.

\section{Results}\label{sec:results}
\subsection{Phase diagram in the absence of the potential barrier}\label{sec: phase diagram}

In the case of $U(x)=0$ and $J=0$, the property of the ground state was studied in Ref.~\onlinecite{Sepulveda2008}; the ground state is solid for $g>g_{\rm c}=21.05\cdots$ and liquid for $g<g_{\rm c}$ and the liquid-solid transition at $g=g_{\rm c}$ is continuous.

\begin{figure}
\includegraphics[width=7.0cm,clip]{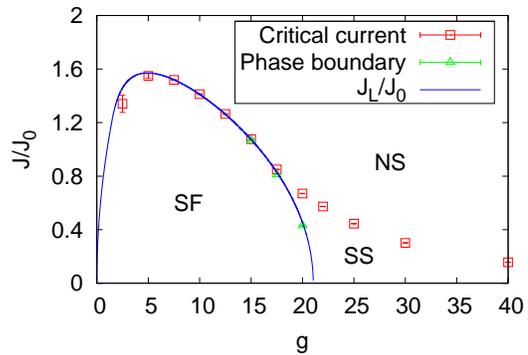}%
\caption{(Color online) Phase diagram in the case of $U(x)=0$ in the plane of $J$ and $g$. SF (SS) denotes the superfluid (supersolid) phase. In the region with $J$ above the critical current (open square), no steady state exists(NS). The solid curve represents the Landau critical current in the liquid phase. Open triangles represent the liquid-solid phase boundary determined from the Bragg peak.}
\label{fig:phasediagram}
\end{figure}
\begin{figure}
\includegraphics[width=7.5cm,clip]{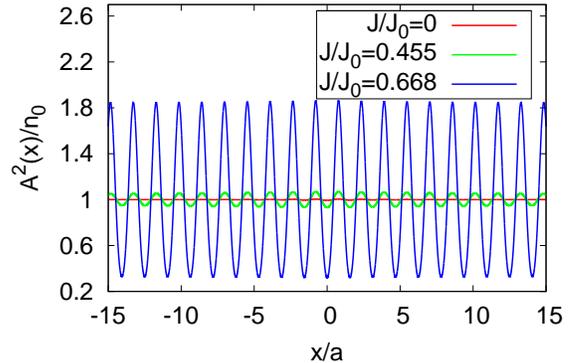}%
\caption{(Color online) Density profile for $U_0/\epsilon_0=0.01$ and $g=20 (<g_{\rm c})$, where $J_{\rm c}/J_0=0.668$. }
\label{fig:g<gcdensity}
\end{figure}

Figure~\ref{fig:phasediagram} shows the phase diagram in the case of $U(x)=0$ and $J\ge 0$. The vertical (horizontal) axis is the current $J$ (the interaction strength $g$). There are three regimes: liquid (superfluid) phase denoted by (SF), supersolid phase (SS), and non-stationary state (NS). The definitions of these regions are as follows. SF is the phase in which the lowest energy state for given $J$ is $|\Psi(x)|/n_0=1$. SS is the phase in which the lowest energy state for given $J$ exhibits a periodic density modulation. NS is the state in which no steady flow solution of eq. (\ref{eq:GPequationforAphiform}) exists for given $J$. Thus, we can not define the phase in this region. Open squares represent the critical current. 
The liquid phase is the region where $g<g_{\rm c}$ and $J<J_{\rm L}$. Here 
\(J_{\rm L}=J_{\rm L}(g)\) denotes the "Landau critical current", which is defined by \cite{Landau1941} 
\begin{eqnarray}
J_{\rm L}=n_0\min\left(\frac{\epsilon_k}{\hbar k}\right)\label{eq:landaucriticalcurrent}
\end{eqnarray}
in terms of the excitation energy $\epsilon_k$ in the liquid phase with $J=0$. 
In the present model, $\epsilon_k$ is obtained analytically as \cite{Sepulveda2008}
\begin{eqnarray}
\epsilon_k=\sqrt{\frac{\hbar^2k^2}{2m}\left[\frac{\hbar^2k^2}{2m}+4n_0V_0\frac{\sin ka}{k}\right]}\label{eq:bogoliubovspectram},
\end{eqnarray}
with which the boundary ($J=J_{\rm L}$) of the liquid phase is determined as shown by the blue curve. 
Open triangles represent the liquid-solid phase boundary determined from the Bragg peak. Those triangles are on the blue curve within the numerical accuracy. 
In Fig.~\ref{fig:phasediagram}, we see that for $g\in [17.5, g_{\rm c}]$, the phase changes from the liquid phase to the solid phase with $J$ increasing.
 This transition corresponds to the current-driven transition between the liquid phase and the modulation phase discussed in Refs.~\onlinecite{Iordanskii1980}, \onlinecite{Pitaevskii1984}, and \onlinecite{Ancilotto2005}.
In the modulation phase, the density has the modulation with the period equal to the inverse of the roton momentum\cite{Pitaevskii1984}. This phase has been discussed to be non-superfluid in Ref.~\onlinecite{Iordanskii1980} and superfluid in Refs.~\onlinecite{Pitaevskii1984} and \onlinecite{Ancilotto2005}. In one-dimensional systems, the modulation phase and solid phase belong to a same phase, contrary to the systems in higher dimensions. 
Figure~\ref{fig:g<gcdensity} illustrates how the density profile 
for $g=20(<g_{\rm c})$ changes with $J$ increasing. 
We confirm that the period of modulation coincides with the inverse of roton momentum and the Bragg peak starts to grow from zero at the phase boundary. 

In the excitation energy (\ref{eq:bogoliubovspectram}), we find that the roton minimum exists only for $g\in [g_{\rm r}(\equiv 9.46\cdots),g_{\rm c}]$ as noted in Ref. \onlinecite{Pomeau1994}. We infer that the Landau critical current yields the liquid-solid phase boundary for $g\ge g_{\rm r}$ while it gives the boundary between the liquid phase and non-stationary state for $g\le g_{\rm r}$. Actually, the curves representing the critical current and the Landau critical current collapse for $g\in [g_{\rm r},17.5]$ as shown in Fig.~\ref{fig:phasediagram}. We cannot confirm our expectation numerically.  
\begin{figure}
\begin{center}
\includegraphics[width=7.5cm,clip]{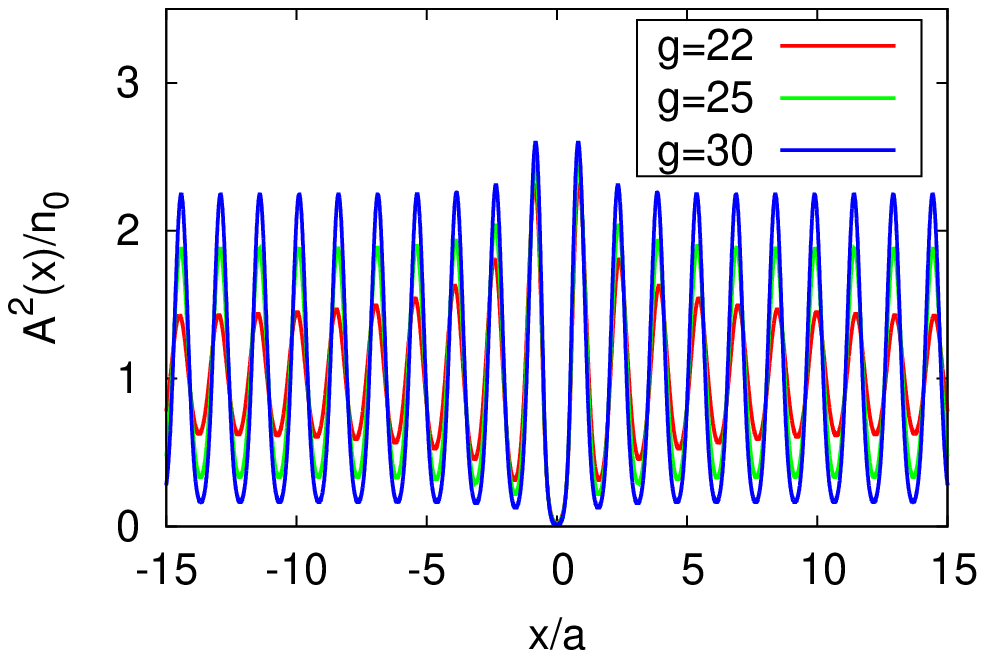}%

\includegraphics[width=7.5cm,clip]{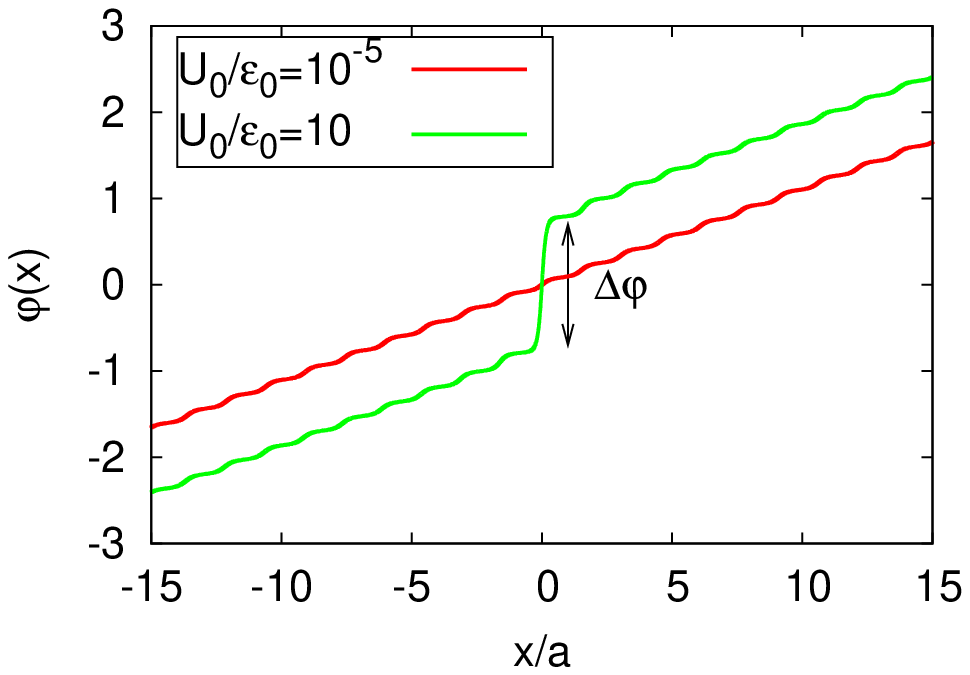}%
\caption{(Color online) (Upper panel) density profile for $(g, J/J_0)=(22, 0.0921),(25, 0.0782),(30, 0.0598)$. For all curves, we set $U_0/\epsilon_0=10$. (Lower panel) the profile of the phase of condensate wave function for $(g, J/J_0)=(25, 0.0782)$ and $U_0/\epsilon_0=10^{-5},U_0/\epsilon_0=10$. $\Delta\varphi$ denotes the phase shift induced by the barrier. }
\label{fig:density}
\end{center}
\end{figure}

\subsection{Josephson effect}\label{sec: Josephson}

Next we consider the case with a potential barrier. 
By numerical calculation, we find that a steady flow solution exists for positive $g$ in the presence of potential barrier. 
%
Figure~\ref{fig:density} shows our numerical results on the spatial dependences of the density of the condensate (upper panel) and the phase of the condensate wavefunction (lower one) in the solid phase ($g>g_{\rm c}$). As seen in the upper panel, the amplitude of modulation becomes larger when $g$ increases in the solid phase. In the lower panel, the gradients of the curves correspond to the local velocities of the condensate. Small wiggle of $\varphi(x)$ has the same period as that of $A(x)^2$. This is a consequence of the spatially uniformity of the current (\ref{continuouseq}). When we compare the two curves in the lower panel, we see that the presence of the barrier affects the phase of the condensate wave function only through the phase shift $\Delta\varphi$ near $x=0$. Obviously this is analogous to the DC Josephson effect in superconductors\cite{Josephson1962}. 

\begin{figure}
\includegraphics[width=7.5cm,clip]{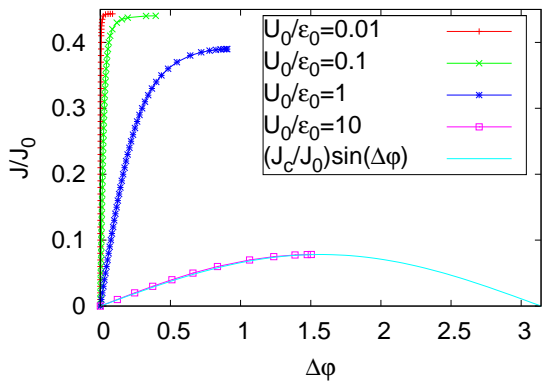}%
\caption{(Color online) $J$-$\Delta\varphi$ characteristic (the Josephson relation) for 
various value of $U_0$ and for $g=25>g_{\rm c}$. For a technical reason, we take $A(x)$ for $U_0/\epsilon_0=10^{-5}$ as $A_0(x)$ in (\ref{eq:phasedifferecnceformula}).}
\label{fig:Josephsonrelation}
\end{figure}

For convenience, the phase-shift $\Delta\varphi$ is quantified as
\begin{eqnarray}
\Delta\varphi&\equiv &\frac{mJ}{\hbar}\int^{L/2}_{-L/2}dx\left[\frac{1}{A(x)^2}-\frac{1}{A_0(x)^2}\right]\label{eq:phasedifferecnceformula}\\
A_0(x)&\equiv &\lim_{U_0\to 0}A(x),
\end{eqnarray}
which is a generalization of the corresponding expressions in Refs. \onlinecite{Baratoff1970} and \onlinecite{Danshita2006} for the liquid phase with superflow. 
Figure~\ref{fig:Josephsonrelation} shows the relation between $J$ and $\Delta\varphi$. In the low barrier limit, the $J$-$\Delta\varphi$ curve becomes steep for small $\Delta\varphi$. In the high barrier limit, $J$-$\Delta\varphi$ curves approach the form
\begin{eqnarray}
J=J_{\rm c}\sin(\Delta\varphi),
\end{eqnarray}
which has the same form as the Josephson relation in superconductors\cite{Josephson1962}. Comparing Fig.~\ref{fig:Josephsonrelation} with the Josephson relation in Fig.~2 of Ref.~\onlinecite{Baratoff1970}, one can find that the results are very similar with each other for $\Delta\varphi\in[0,\Delta\varphi_{\rm c}]$ with $\Delta\varphi_{\rm c}$ defined by the phase shift at the critical current.
\begin{figure}
\includegraphics[width=7.0cm,clip]{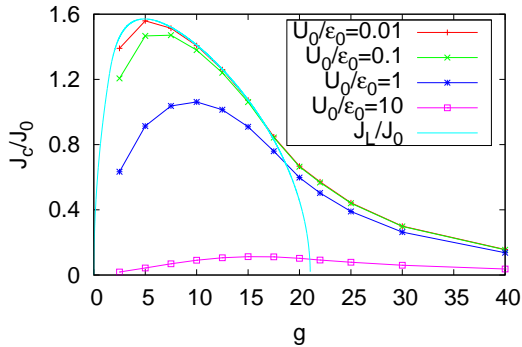}%
\caption{(Color online) $g$-dependence of the critical current $J_{\rm c}$ for various strength of potential barrier. The solid curve represents the Landau critical current. }
\label{fig:phasediagramwithU}
\end{figure}
Figure \ref{fig:phasediagramwithU} shows $g$-dependence of the critical current for the Josephson effect. We see that the current-driven solid state at $g<g_{\rm c}$ can yield the Josephson effect. In this sense, the superflow property of current-driven solid state remains even in the presence of an obstacle. 

In Fig.~\ref{fig:phasediagramwithU}, we notice that the $J=J_{\rm c}(g)$ curve is not affected by the presence of the curve representing the Landau critical current for $g\in [17.5,g_{\rm c}]$. 
We thus expect that superflow property does not change so much around this boundary for $g\in [17.5,g_{\rm c}]$. 

\subsection{persistent current in a static ring}\label{sec: persistent}

We consider the case where the boundary condition (\ref{eq: bc-Psi}) is imposed. This case describes the condensate forming a ring with perimeter $L$. The boundary condition (\ref{eq: bc-Psi}) is rewritten as (\ref{eq: bc-A}) and
\begin{equation}
\varphi(x+L)=\varphi(x)+2\pi M,
\label{eq: bc-varphi}
\end{equation}
with an integer $M$ (a winding number). Among the stationary solutions obtained in \ref{sec: Josephson}, realized are only the solutions specified by the discrete set of $J$ satisfying 
\begin{eqnarray}
\frac{mJ}{\hbar}\int_{-L/2}^{L/2}dy\frac{1}{A(y)^2}=2\pi M,
\label{eq: varphi-J-M}
\end{eqnarray}
which immediately follows from (\ref{eq: varphi-A}) and (\ref{eq: bc-varphi}). 
The solution specified by a nonzero $M$, which carries a nonzero $J$, is metastable because the winding number $M$ cannot change continuously\cite{Leggett2006note}; the only way for a current carrying state to relax to another state specified by a winding number $M'(\ne M)$ is to cause a phase slip\cite{Anderson1966}. When a phase slip occurs, the system undergoes an intermediate state with a zero point $x_0$ of condensate wave function $\Psi(x)$ such that $\Psi(x_0)=0$ and the phase $\varphi(x_0)$ cannot be defined. Such an intermediate state has energy higher than the initial state by the local depletion of condensation energy in the vicinity of $x=x_0$. The metastable state with nonzero $M$ (i.e. nonzero $J$) yields a persistent current if the energy barrier that the metastable state has to overcome to relax to another state is sufficiently large and the relaxation time is much larger than other time scales. 

\subsection{Hess-Fairbank effect, nonclassical rotational inertia and superfluid fraction in a moving ring}\label{sec: Hess}

\begin{figure}
\includegraphics[width=7.0cm,clip]{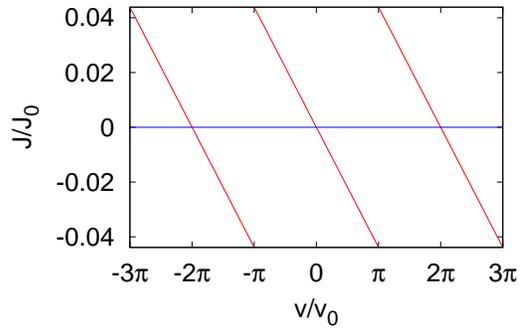}%
\caption{(Color online) $v$-dependence of $J(v)$ for $g=25$ and $d=a$, $U_0=\epsilon_0$. Here, $v_{0}\equiv \hbar/mL$.}
\label{fig: J-v}
\end{figure}

\begin{figure}
\includegraphics[width=7.0cm,clip]{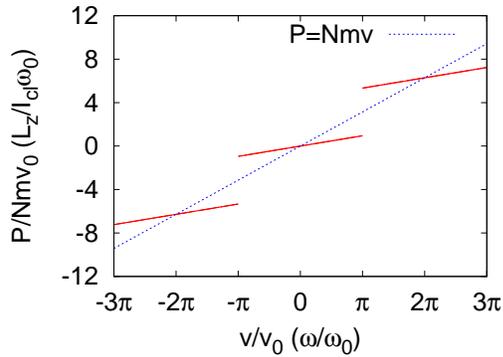}%
\caption{(Color online) $v$-dependence of total momentum $P$ for $g=25$ and $d=a$, $U_0=\epsilon_0$. Alternatively, this curve can be regarded as the relation between angular momentum $L_z=RP$ of the condensate and angular velocity $\omega=v/R$ of the container with the annular geometry with radius $R=L/(2\pi)$. The unit of angular velocity is $\omega_0=\hbar/(mLR)$.}
\label{fig: P-v}
\end{figure}

On the basis of results in the previous section, we discuss supersolidity in the presence of an obstacle moving with a constant velocity $v$ under the periodic boundary condition (\ref{eq: bc-Psi}). This situation describes the condensate in a toroidal geometry with radius $R=L/(2\pi)$ and sufficiently small cross section. The equation under a moving obstacle  
\begin{eqnarray}
&&i\hbar\frac{\partial \Psi(x,t)}{\partial t}=-\frac{\hbar^2}{2m}\frac{\partial^2}{\partial x^2}\Psi(x,t)+\left[U(x-vt)-\mu\right]\Psi(x,t)\nonumber \\
&&\quad \quad +\int^{L/2}_{-L/2}dy V(x-y)|\Psi(y,t)|^2\Psi(x,t),
\label{eq:t-dep-GPequationforpsiform-moving}
\end{eqnarray}
is rewritten as
\begin{eqnarray}
&&i\hbar\frac{\partial \Psi'(x',t)}{\partial t}\Big|_{x'}\nonumber\\
&=&-\frac{\hbar^2}{2m}\frac{\partial^2\Psi'}{\partial x'^2}\Big|_{t}+i\hbar v\frac{\partial\Psi'}{\partial x'}\Big|_{t}
+\left[U(x')-\mu\right]\Psi'(x',t)\nonumber \\
&&\quad \quad +\int^{L/2}_{-L/2}dy' V(x'-y')|\Psi'(y',t)|^2\Psi'(x',t),
\label{eq:t-dep-GPequationforpsiform-moving-Psi-prime} 
\end{eqnarray}
in terms of $x'=x-vt$ and $\Psi'(x',t)=\Psi(x,t)$. In the moving obstacle, the stable state (we call this state the ground state in the moving frame) yields the minimum value of the functional\cite{NozieresPines,Leggett2006}
\begin{equation}
\tilde{E}[\Psi',\Psi'^*]=E[\Psi',\Psi'^*]- v P[\Psi',\Psi'^*],
\label{eq: tilde E}
\end{equation}
under the constraint (\ref{eq: chemical}) and the periodic boundary condition 
(\ref{eq: bc-Psi}). In (\ref{eq: tilde E}), $P[\Psi',\Psi'^*]$ denotes the total momentum 
\begin{eqnarray}
&&P[\Psi',\Psi'^*]\nonumber\\
&&=\frac{\hbar}{2i}\int_{-L/2}^{L/2}\hspace{-0.9em}dx'\hspace{-0.25em}\left[\Psi'^*(x',t)\frac{\partial \Psi'(x',t)}{\partial x'}-\Psi'(x',t)\frac{\partial \Psi'^*(x',t)}{\partial x'}\right], \nonumber\\
\label{eq: P}
\end{eqnarray}
which is nothing but $m\int_{-L/2}^{L/2}J(x,t)d x$ with (\ref{eq: Jxt}).
The ground state in the moving frame is a stationary solution of (\ref{eq:t-dep-GPequationforpsiform-moving-Psi-prime}) because 
(\ref{eq:t-dep-GPequationforpsiform-moving-Psi-prime}) is written as
\begin{eqnarray}
&&i\hbar\frac{\partial \Psi'(x',t)}{\partial t}=\frac{\delta\tilde{E}[\Psi',\Psi'^*]}{\delta \Psi'^*}-\mu\frac{\delta\int_{-L/2}^{L/2}|\Psi(x',t)|^2dx'}{\delta\Psi'^*}.\nonumber\\
\label{eq:t-dep-GPequationforpsiform-moving2} 
\end{eqnarray}
Let the stationary solution to (\ref{eq:t-dep-GPequationforpsiform}) for a given $J\in [-J_{\rm c},J_{\rm c}]$ obtained in the previous section be denoted by $\Psi_J$ and the accompanying chemical potential $\mu_J$. The stationary solution to (\ref{eq:t-dep-GPequationforpsiform-moving2}) is then given by
\begin{equation}
\Psi'(x')=\Psi_J(x')e^{i mv x'/\hbar},\label{eq: Psi-PsiJ}
\end{equation}
with $\mu=\mu_J-mv^2/2$. Note that from the boundary condition (\ref{eq: bc-Psi}), it follows that
\begin{equation}
\Psi_J(x'+L)=\Psi_J(x')e^{-i mv L/\hbar}.\label{eq: bc-PsiJ}
\end{equation}
Substituting (\ref{eq: Psi-PsiJ}) into (\ref{eq: tilde E}), we obtain
\begin{equation}
\tilde{E}[\Psi',\Psi'^*]=E[\Psi_J,\Psi_J^*]+\frac{Nmv^2}{2}.
\end{equation}
$E[\Psi_J,\Psi_J^*]$ is even with respect to $J$ and increases with $J$ increasing. 
Thus what we seek is $J$ with minimum $|J|$ such that the boundary condition 
(\ref{eq: bc-PsiJ}) is satisfied. From (\ref{eq: varphi-A}), 
\begin{equation}
\varphi(L/2)-\varphi(-L/2)=\frac{m J}{\hbar}\int_{-L/2}^{L/2}dy\frac{1}{A_J(y)^2}
\end{equation}
with $A_J(y)\equiv|\Psi_J(y)|$ follows. From  this and (\ref{eq: bc-PsiJ}), we obtain 
\begin{equation}
v=-\frac{J}{L}\int_{-L/2}^{L/2}dy \frac{1}{A_J(y)^2}+\frac{2\pi \hbar M}{mL}
\label{eq: v-J}
\end{equation}
with an integer $M$.  Among the solutions $J$ to (\ref{eq: v-J}) under a given $v$, let denote by $J(v)$ the solution with minimum $|J|$. Using the results on $A_J$, we can obtain $J(v)$ as shown in Fig.~\ref{fig: J-v}. $J(v)$ is periodic with respect to $v$ because $v$ and $v+2\pi\hbar/(mL)$ yield the same boundary condition (\ref{eq: bc-PsiJ}) on $\Psi_J$.
Using this result, we can derive the $v$-dependence of the total momentum,
\begin{eqnarray}
P&=&\frac{\hbar}{2i}\int_{-L/2}^{L/2}\hspace{-0.9em}dx\left[\Psi^*(x,t)\frac{\partial \Psi(x,t)}{\partial x}-\Psi(x,t)\frac{\partial \Psi^*(x,t)}{\partial x}\right]\nonumber\\
&=&\frac{\hbar}{2i}\int_{-L/2}^{L/2}dx'\left[\Psi'^*(x')\frac{d\Psi'(x')}{d x'}-\Psi'(x')\frac{d\Psi'^*(x')}{dx'}\right]\nonumber\\
&=&m\left[v\int_{-L/2}^{L/2} dx|\Psi_J(x')|^2 +J L\right]=Nmv+m JL.\nonumber\\
\end{eqnarray}
With input shown in Fig.~\ref{fig: J-v}, we plot $P$ as a function of $v$ as shown in Fig.~\ref{fig: P-v}. The dotted line shows the line $P=Nmv$, which represent the normal state (i.e. the state does not exhibit superflow). When $|v|\le v_0\equiv \hbar/(mL) $, the slope of the $P/(Nmv)-v/v _0$ curve represents $1-f_{\rm s}=f_{\rm n}$. Here $f_{\rm s}\equiv (Nmv-P)/Nmv$ and $f_{\rm n}$ represent, respectively, the superfluid and non-superfluid fractions. Note that Figure~\ref{fig: P-v} can be regarded as a normalized plot of the relation between angular momentum $L_z=RP$ and angular velocity $\omega=v/R$ of the rotating container (ring) with the radius $R=L/(2\pi)$. Compare this figure with e.g. Fig.~3.2 of Ref.~\onlinecite{Leggett2006}. 
Here $\omega_0=v_0/R=\hbar/(mLR)$ and $I_{\rm cl}=Nm R^2$ (classical rotational inertia). In  this regard, Figure~\ref{fig: P-v} shows that the supersolid state exhibits nonclassical rotation inertia $I=I_{\rm cl}f_{\rm s}$, which is a sign of the Hess-Fairbank effect in supersolid in the circular asymmetric container. 

\section{Discussion}\label{sec:discussion}


In this paper, we studied one-dimensional supersolids in the framework of the mean-field theory. In low-dimensional systems, quantum fluctuations are large and destroy long-range order\cite{Mermin1966,Hohenberg1967}. Therefore, mean-field theories is not necessarily applicable in low dimensional systems. Roughly speaking, our mean-field description is valid to describe phenomena over the time scale shorter than the inverse of nucleation rates of topological defects (kink);  Those defects destroy the phase coherence or positional order in an ordered state. Importance of the results in the present paper lies in simplicity of models showing the Josephson effect in supersolid phase and implication to systems in higher dimensions; Our results strongly suggest that the Josephson effect is likely to occur in supersolids in higher dimensions. 
%

In an earlier work\cite{Pomeau1994}, on the other hand, it was concluded that non-dissipative mass flow around an obstacle is impossible in a two-dimensional supersolid. There are two possibilities of the cause of the difference. One is the way of setting boundary conditions. In Ref.~\onlinecite{Pomeau1994}, the details of the boundary conditions are not available. The boundary conditions may determine whether or not dissipationless flow is possible. The other possibility is due to the difference of shapes of the obstacle. Although two-dimensional supersolid could exhibit the Josephson effect, dissipationless flow circumventing the obstacle may be impossible. A two-dimensional solid is characterized by positional order and orientational order( see e.g. Refs.~\onlinecite{Nelson} and \onlinecite{Strandburg}). A plate-shaped obstacle corresponding to the Josephson effect may distort only positional order, whereas a disk-shaped obstacle may distort positional order and orientational order.
A future problem is to understand these two results, which seemingly contradicts with each other, in a consistent way. 

We now discuss how the phase diagram shown in Fig.~\ref{fig:phasediagram} will change in other spatial dimensions where the two-body interaction $V(r)\propto\theta(a-r)$ with a range $a(>0)$ is the same type as 
that in Eq.~(\ref{eq:GPequationforpsiform}). 
As in the one-dimensional model, the excitation energy $\epsilon_k$ in two or three dimensions has the roton minimum when $g$ is larger than a certain value $g_{\rm r}$ and the Landau critical current $J_{\rm L}$ becomes zero at a coupling constant $g_{\rm c} (>g_{\rm r})$.  
In two or three dimensions, the first-order phase transition between the liquid phase and solid phase at $J=0$ was considered to occur at a coupling constant $g=g_{1}\in (g_{\rm r},g_{\rm c})$\cite{Pomeau1994}. At sufficiently small $J$, the liquid-solid transition is expected to be first-order.
The phase boundary between the liquid phase and a solid phase is represented by a curve that starts from $(J,g)=(0,g_1)$ and terminates when it meets the curve $(J,g)=(J_{\rm L}(g),g)$ representing the Landau critical current. 
Although the modulation phase (i.e., the state with laminar pattern of density)  in two and three dimensions might be no longer stable, there still remains a possibility that this phase is metastable and contributes as an intermediate state to a transient dynamics relaxing from a liquid state to a solid state with $J>0$. 

We also discuss how the critical current $J_{\rm c}(g)$ of the Josephson effect looks like in two and three dimensions. We expect the overall feature of the curve $J=J_{\rm c}(g)$ in those dimensions to be similar to that shown in Fig.~\ref{fig:phasediagramwithU}. As a notable difference between one dimension and higher dimensions, we expect that the curve $J=J_{\rm c}(g)$ has a cusp or kink when it crosses the first-order phase boundary between the liquid and the solid phases. 

From Fig.~\ref{fig:density} (lower panel) and Fig.~\ref{fig:Josephsonrelation}, the superflow property of the solid phase seems to be similar with that of the conventional superfluid. We discuss a possible difference between the two phases. In conventional Josephson junctions, $J$-$\Delta\varphi$ characteristics has a curve connecting the points $(J,\Delta\varphi)=(J_{\rm c},\Delta\varphi_{\rm c})$ and $(J,\Delta\varphi)=(0,\pi)$. This curve corresponds to the unstable steady state.  In Ref.~\onlinecite{Hakim1997}, it was shown for GP equation with contact-type two-body interaction and the potential barrier $U(x)\propto \delta(x)$, there exists only a single branch that corresponds to the unstable steady state. In the present case we study, on the other hand, the number of unstable and metastable branches are unknown. The structure (i.e., the number and stabilities) of the branches corresponding to steady states in the solid phase might be different from that in the conventional superfluid. From the context of non-linear physics, this implies that the critical current state in the solid phase in the present model might be categorized as a different type of bifurcation points from the critical current state in the conventional superfluids. 
The bifurcation theory discusses parameter-dependence of the number and stability of stationary solutions of time-dependent non-linear differential equations\cite{Guckenheimer1983}. The bifurcation point means the critical parameter at which the number or stability of stationary solutions changes. In the present case, equation~(\ref{eq: relaxation}) has at least one stationary stable solution for $J<J_{\rm c}(g)$ and no stationary solutions for $J>J_{\rm c}(g)$. The critical parameter $J=J_{\rm c}(g)$ is thus a bifurcation point. It is important to identify the type of bifurcation points in the present model in the sense that the type of the bifurcation in the critical current state governs the dynamics in the breakdown of superfluids, such as the emission rate of the topological defects (solitons and vortices) around obstacles at $J$ slightly larger than the critical value\cite{pomeau93,huepe00,rica01,pham02}. 
If the critical current state in the solid phase in the present model is identified with a different type of bifurcation point from that in conventional superfluids, then the dynamical properties in breakdown process of superfluid properties is also expected to be unconventional. The dynamics of the solid phase above the critical current and identification of the bifurcation type are addressed in a future study.  

\section{Conclusion}
\label{sec:conclusion}
In conclusion, we showed that superflow properties (Josephson effect, Persistent current and Hess-Fairbank effect) of a solid phase are maintained even in the presence of an obstacle in a one-dimensional GP model with finite-range interaction. 
A generalization to higher dimensions and the study of the mechanism of breakdown of supersolidity are important future issues.  

We thank S.~Watabe, D.~Takahashi, T.~Minoguchi, M.~Kobayashi, S.~Tsuchiya, I.~Danshita, D.~Yamamoto, M.~Takahashi, C.~Josserand and F.~Ancilott for useful discussions. M. K. acknowledges support by Grant-in-Aid for JSPS Fellows (239376).
This work is supported by KAKENHI (21540352) from JSPS and (20029007) from MEXT in Japan. 

\appendix
\section{}
\label{sec:appendixA}

In this appendix, we show that the stationary condition of $K$ with respect to the variation $A(x)\rightarrow A(x)+\delta A(x)$ and $\varphi(x)\rightarrow \varphi(x)+\delta \varphi(x)$ under the constraint (\ref{eq: chemical}) and the boundary condition (\ref{eq: bc-A}). Following the standard procedure of variation under a constraint, we consider the variation of the functional $\tilde{K}=K-\mu \int_{-L/2}^{L/2}dx|\Psi(x)|^2 $. For the functional $$\tilde{E}=E-\mu \int_{-L/2}^{L/2}dx|\Psi(x)|^2 ,$$ we obtain
\begin{eqnarray}
&&\tilde{E}[A(x)\rightarrow A(x)+\delta A(x),\varphi(x)\rightarrow \varphi(x)+\delta \varphi(x)]\nonumber\\
&&-\tilde{E}[A(x),\varphi(x)]\nonumber\\
&&=\delta \tilde{E}+O((\delta A(x))^2,\delta A(x)\delta \varphi(x),(\delta \varphi(x))^2), 
\end{eqnarray} 
with
\begin{eqnarray}
\delta \tilde{E}&=&\int_{-L/2}^{L/2}dx\delta A(x){\cal G}[A(x),\varphi(x)]\nonumber\\
&+&\int_{-L/2}^{L/2}dx\delta\varphi(x)\frac{d}{dx}\left[-\frac{\hbar^2}{m}A(x)^2\frac{d\varphi(x)}{d x}\right]\nonumber\\
&+&\frac{\hbar^2}{m}\left[\frac{dA(x)}{dx}\delta A(x)\right]_{x=-L/2}^{x=L/2}\nonumber \\
&+&\frac{\hbar^2}{m}\left[A(x)^2 \frac{d\varphi(x)}{dx}\delta \varphi(x)\right]_{x=-L/2}^{x=L/2}.\nonumber\\
\label{eq:deltaE}
\end{eqnarray} 
The first and second terms in RHS vanish when $A(x)$ and $\varphi(x)$ satisfy (\ref{eq:GPequationforamplitude1}) and (\ref{eq:contiunuityequationfora}). The third term in RHS of (\ref{eq:deltaE}) becomes zero from the boundary condition (\ref{eq: bc-A}). The last term becomes
\begin{eqnarray}
\frac{\hbar^2}{m}\left[A(x)^2 \frac{d\varphi(x)}{dx}\delta \varphi(x)\right]_{x=-L/2}^{x=L/2}\nonumber \\
=\hbar J\left[\delta\varphi(L/2)-\delta\varphi(-L/2)\right].
\end{eqnarray}
It thus follows that 
\begin{eqnarray}
\delta \tilde{K}=\delta E-\hbar J\left[\delta\varphi(L/2)-\delta\varphi(-L/2)\right]=0.
\end{eqnarray}

\section{}
\label{sec:appendixB}

\begin{figure}
\includegraphics[width=7.0cm,clip]{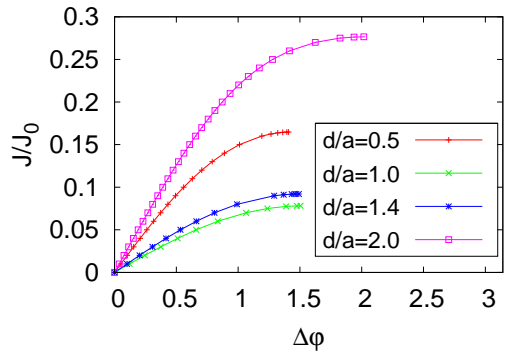}%
\caption{(Color online)  $d$-dependence of the Josephson relation for $U_0/\epsilon_0=10$ and $g=25$.}
\label{fig:josephson_d-dep}
\end{figure}

\begin{figure}
\includegraphics[width=7.0cm,clip]{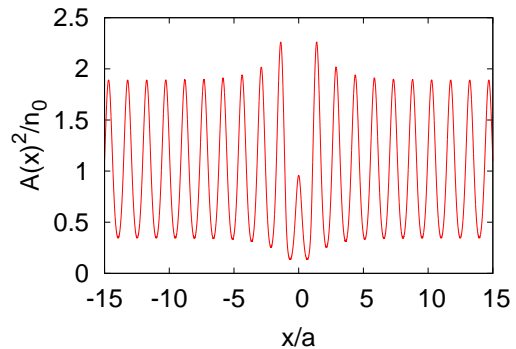}%
\caption{(Color online)  Density profile for $g=25, J=0, U_0=10\epsilon_0$, and $d=2a$.}
\label{fig:d=2_dependence}
\end{figure}

In this appendix, we discuss the $d$-dependence of our results ($d$ denotes the range of potential barrier defined in (\ref{eq: U-def}). Figure \ref{fig:josephson_d-dep} shows that the Josephson relation for various value of $d$. We find that $d$-dependence of the critical current is nonmonotonic in contrast to $U_0$-dependence. This behavior of the critical current is due to the commensurability of the width of the potential barrier and the period of density modulations. Fig.~\ref{fig:d=2_dependence} shows that the density profile for $d/a=2$, $U_0/\epsilon_0=10$, $J=0$, and $g=25$. There are two local minima of the density in the potential barrier. This result is different from that of $d\le1.4a$. However, Josephson effect occurs for $d=2a$ as shown in Fig. \ref{fig:josephson_d-dep}. This result suggests that Josephson effect in one-dimensional supersolids occurs regardless of the width of the obstacle.


\begin{thebibliography}{99}
\bibitem{note-1d}
Note that $[\Psi(x)]={\rm L}^{-1/2}$ and $[J]={\rm T}^{-1}$ because we consider a one-dimensional model.
\end{thebibliography}

\end{document}